\documentclass{aa}  

\usepackage{graphicx}
\usepackage{txfonts}
\usepackage{lipsum}
\usepackage{subcaption}         
\usepackage{lscape}             
\usepackage{placeins}           

\usepackage{orcidlink}
\newcommand{\orcid}[1]{\orcidlink{#1}}
                                
\usepackage[]{hyperref}

\begin{document}

   \title{The Connection Between the Growth of Massive Black Holes and Their Dynamical Histories}

%

   \author{Lee A. ~Whitaker\inst{1}\fnmsep\thanks{Corresponding author: leewhitaker@umail.ucc.ie} ~\orcid{0009-0001-0316-9341}
        \and Michael ~Tremmel\inst{1} ~\orcid{0000-0002-4353-0306}
        \and Angelo ~Ricarte\inst{2} ~\orcid{0000-0001-5287-0452}
        \and Priyamvada ~Natarajan\inst{3} ~\orcid{0000-0002-5554-8896}
        }

   \institute{School of Physics, University College Cork, Cork, Ireland T12 K8AF
   \and Center for Astrophysics | Harvard \& Smithsonian, 60 Garden Street, Cambridge, MA 02138, USA
   \and Department of Astronomy, Yale University, New Haven, CT 06511, USA}

   \date{Received Month XX, 20XX}

 
  \abstract
   {Massive black holes (MBHs) grow through both gas accretion and mergers and it remains unclear how variations to MBH growth and dynamical histories contribute to the spread in empirical scaling relations between MBHs and their host galaxies.}
   {We examine the ex- and in-situ growth of MBHs within large-scale cosmological simulations and compare the evolution of an outlier population of under-grown MBHs to the general population.} 
   {We use the Romulus25 uniform volume cosmological simulation and the RomulusC cluster zoom-in simulation to examine the growth and merger histories of MBHs. We identify central MBHs that are undergrown relative to other MBHs in similar mass galaxies and investigate their accretion, merger, and orbital histories within the simulations.}
   {The contribution of ex-situ growth increases with MBH mass, with mergers becoming increasingly important in the most massive galaxies. We find no difference between cluster and field environments in terms of the importance of ex-situ growth, which is generally very low. Under-grown MBHs have experienced fewer mergers and have had complex dynamical histories, spending significant time `drifting' within the galaxy or`competing' with another MBH. Other under-grown MBHs have experienced a `late entrance' into a galaxy which until then had no central MBH.}
   {Our results demonstrate that MBH assembly is dominated by in-situ growth, which in turn is impacted by the dynamical histories of MBHs within assembling galaxies. Prolonged off-centre evolution can suppress accretion and contribute to the formation of under-grown central MBHs at $z=0$.}

   \keywords{galaxies:evolution -- 
            black hole physics -- 
            galaxies:agn
               }

   \maketitle

\nolinenumbers

\section{Introduction}

    Massive black holes (MBHs) are a fundamental, yet still not fully understood, component of galaxy evolution. These black holes, with masses of $10^5-10^{10}$ M$_{\odot}$, are found at the centres of most galaxies, from massive ellipticals to bulgeless disc galaxies \citep[e.g.][]{kormendy_inward_1995,kormendy_coevolution_2013,reines_relations_2015,reines_dwarf_2013,moran_black_2014}. Empirical scaling relations between MBH mass and host galaxy properties such as stellar mass and velocity dispersion \citep[e.g.][]{haring_black_2004,kormendy_coevolution_2013,schramm_black_2013}, together with theoretical models of MBH assembly in a cosmological context \citep{natarajan_mass_2012} point to a co-evolutionary connection between black holes and galaxies. However, the mechanisms by which these black holes assemble their mass remains an open question. Thus, it becomes interesting to examine more closely departures from these scaling relations, i.e. MBHs which are larger or smaller than typical given their host galaxy properties.

    Observationally, as we attain more data on MBHs in lower mass galaxies and MBHs at low redshift, deviations from typical scaling relations become apparent, potentially hinting at multiple relations that reflect different evolutionary states \citep[e.g.][]{schulze_accounting_2014,reines_relations_2015, greene_2020}.
     Variability in accretion histories, merger-driven growth, and environmental effects could all play a role in setting MBH masses and in producing the observed scatter about the scaling relations. It is therefore interesting to consider how these deviations could provide insight into the otherwise hidden growth and dynamical histories of MBHs and how this relates to the balance between in-situ vs ex-situ growth. 

     One uncertainty when examining observations is in the past merger history of the MBHs. In-situ growth through direct gas accretion has been studied extensively, both theoretically and observationally \citep[e.g.][]{soltan_masses_1982,hopkins_how_2010,alexander_what_2012}, and it is often what is invoked at high redshift to understand the origins of massive black holes \citep[e.g.][]{larson_jwst_2023}. Empirical relations between MBH accretion rate and galaxy star formation rate \citep{mullaney_agnms_2012} or stellar mass \citep{yang_linking_2018}, indicate that the origins of empirical relations can be largely explained by connecting a MBH's ability to grow to the gas supplied by its host galaxy. In contrast, ex-situ growth - mass acquired through the mergers of pre-existing black holes brought in by galaxy mergers -- can be more difficult to disentangle obsesrvationally. Early work by \citet{volonteri_assembly_2003} explored the contribution of black hole mergers to MBH mass assembly in a hierarchical cosmological framework, and subsequent studies have expanded on these findings, including in the context of MBHs in dwarf galaxies and low-mass hosts \citep[e.g.][]{natarajan_seeds_2014, pacucci_active_2021}. More recently, \citet{porras-valverde_tracking_2026} examined MBH growth channels within semi-analytic models, finding that black hole mergers dominate mass assembly only for the most massive MBHs at late times. Large-scale cosmological simulations also predict that mergers are vitally important to the early population of MBHs \citep{weinberger_supermassive_2018, bhowmick_2025}, indicating that the mass function of black holes at high redshift could retain vital information about their formation properties.
    
    The dynamics of MBHs can also be complex. Early galaxy interactions can bring MBHs together, but they can also perturb them away from galactic centre. The timescale for MBHs to sink to the centres of galaxies following galaxy mergers can be of order Gyrs \citep{bellovary_first_2011,tremmel_2018_mergetime, volonteri_2020, bellovary_2021, chen_df_2022} and can be particularly important in high redshift, low mass galaxies \citep{pfister_2019}. This timescale depends on the properties of the merging galaxies, but can be further complicated by structures on smaller scales, such as gaseous clumps or bars \citep{roskar_2015, bortolas_torquesBH_2020, bortolas_barsBHs_2022}. MBHs that spend significant time away from the centres of galaxies will naturally grow less due to insufficient gas supply. Various modern cosmological simulations that attempt to realistically track the dynamics of MBHs now predict that massive galaxies, which undergo a rich history of minor mergers, host a significant fraction of `wandering' MBHs, which have experienced both less mergers and less accretion than their counterparts in the centres of galaxies \citep{tremmel_2018_wanderers, Ricarte+2021a,Ricarte+2021b,dimatteo_2023_wanderers}.
    
    Cosmological hydrodynamic simulations are useful for investigating the formation and growth of MBHs within a fully self-consistent galaxy formation context. Large-volume simulations \citep[e.g.][]{springel_simulations_2005,dubois_black_2014,schaye_eagle_2015,dave_simba_2019,weinberger_simulating_2017} have yielded important insights into the interaction between MBHs and their host galaxies, including feedback-regulated growth. However, such simulations are limited by finite resolution and the need for sub-grid prescriptions to model MBH growth, accretion, dynamics, and formation. Still, within these constraints, they allow us to track both gas accretion and black hole mergers over cosmic time, and to connect MBH assembly to the hierarchical assembly of their host galaxies. However, if we care about the dynamical and merger histories of MBHs, it is important for simulations to a) seed MBHs in realistic environments at high redshift and b) accurately track their dynamics within and following galaxy mergers.
    
    In this paper, we use the Romulus cosmological simulations \citep{tremmel_romulus_2017, tremmel_romulusc_2019} to quantify the contribution of ex-situ growth to the overall mass assembly of MBHs and to investigate the origins of an under-grown MBH population. Romulus combines relatively high mass and spatial resolution with a sub-grid model for MBH formation that \textit{predicts} where MBHs form at $z>5$, resulting in a naturally evolving MBH occupation fraction \citep{tremmel_2024_of}. Romulus also includes a sub-grid model for dynamical friction that results in realistic sinking timescales \citep{tremmel_wandering_2015}, and therefore more realistic merger histories and dynamics.
    Using Romulus, we identify a population of under-grown MBHs that lie significantly below the $M_{\rm BH}$--$M_\ast$ relation at $z=0$. We then examine this population with respect to the underlying MBH/galaxy population in terms of both their merger histories and dynamical evolution, finding that complex dynamical histories are the root cause of their stunted growth.
    
    The paper is organised as follows. In Section~\ref{sec:simulations} we briefly describe the relevant aspects of the Romulus simulations and our MBH and galaxy samples. In Section\ref{sec:results} we present our results examining a population of under-grown MBHs identified in the simulations.
    We discuss the broader implications of our results in Section~\ref{sec:discussion} and summarise our conclusions in Section~\ref{sec:conclusions}.

\section{The Simulations}
\label{sec:simulations}

    In this work we use the Romulus suite of cosmological hydrodynamic simulations \citep{tremmel_romulus_2017,tremmel_romulusc_2019}, run with the $N$-body + smoothed particle hydrodynamics (SPH) code \textsc{ChaNGa} \citep{menon_changa_2015}. The simulations follow the coupled evolution of dark matter, gas, stars, and massive black holes (MBHs) in a $\Lambda$CDM universe with a flat $\Lambda$CDM cosmology and $\Omega_{m}=0.3086$, $\Omega_{\Lambda}=0.6914$, $h=0.6777$, and $\sigma_8=0.8288$ \citep{planck_planck_2014}.
    
    {\sc Romulus25} is a uniform cosmological box of side length $L_{\rm box} = 25\,{\rm cMpc}$ evolved from $z \sim 99$ to $z = 0$. {\sc RomulusC} is a zoom in simulation \citep{katz_and_white_1993_zoomin} of a galaxy cluster with M$_{200} \sim 10^{14}\,{\rm M}_\odot$ at $z=0$, simulated within a 50 Mpc volume sampled at significantly lower resolution to maintain realistic large-scale torques. Both simulations have mass resolution of $3.39 \times 10^5$ M$_{\odot}$,  $2.12 \times 10^5$ M$_{\odot}$, and $6.36\times10^4$ M$_{\odot}$ for dark matter, gas, and stars respectively. The simulations have a spline kernel gravitational softening of $350\,{\rm pc}$ (equivalent to 250 pc plummer softening)
    
    The sub-grid physics in both simulations includes star formation and stellar feedback \citep{stinson_star_2006}, metal enrichment, low temperature metal cooling \citep{guedes_forming_2011}, a UV background with self-shielding \citep{pontzen_damped_2008}, and radiative primordial gas cooling. In addition, Romulus tracks MBH formation, accretion,and feedback. A key feature of Romulus is the sub-grid treatment of dynamical friction acting on MBHs \citep{tremmel_wandering_2015}, which allows black holes to sink to galaxy centres and merge on physically motivated timescales. This is in contrast to more typical, ad hoc repositioning schemes \citep[e.g.][]{schaye_eagle_2015,sijacki_illustris_2015,weinberger_simulating_2017}, which result in premature merger timescales, often with MBH mergers occurring before even the original host galaxies complete their merger \citep{buttigieg_advect_2025}.
    
    MBH seeds are introduced in gas reaching 30-times the density threshold for star formation ($n > 3\,{\rm cm}^{-3}$), at low-metallicity ($Z < 3 \times 10^{-4}$), and moderately warm temperatures ($9500\,{\rm K} < T < 10,000\,{\rm K}$) with initial masses of $10^6\,{\rm M}_\odot$ \citep{tremmel_romulus_2017}. This is meant to broadly represent regions where gas is likely to collapse into either a supermassive star or a dense nuclear star cluster, two of the main formation mechanisms for massive MBH seeds \citep{volonteri_seeds_2012}. The seed mass, however, is larger than what is physically realistic. This is in part motivated by the fact that, with limited resolution, the MBHs are seeded later than they should be because those early structures are poorly resolved, causing the simulation to miss their initial growth phases. However, there is also the practical necessity that this is the smallest mass for which the dynamics can be realistically tracked with the dynamical friction prescription due to the underlying baryonic and dark matter mass resolution \citep{tremmel_wandering_2015}. Once seeded, MBHs grow through a combination of gas accretion and mergers with other MBHs. Gas accretion is modelled with a modified Bondi–Hoyle prescription that depends on the local gas properties and relative kinematics, including differet prescriptions for rotationally supported gas. MBH accretion is assumed to be Eddington limited. AGN feedback couples a fraction of the accreted rest-mass energy to the surrounding gas as thermal energy. This occurs isotropically, but results in large-scale outflows that can impact the global star formation and circumgalactic medium \citep{tremmel_romulusc_2019, sanchez_2019_AGN_CGM, sanchez_2024}. Mergers between MBHs occur when two MBHs are separated by less than twice the gravitational softening length, $2\epsilon_g$, and have sufficiently low relative velocities to be considered gravitationally bound, satisfying $\frac{1}{2}\Delta\mathbf{v}<\Delta\mathbf{a}\cdot\Delta\mathbf{r}$, where $\Delta\mathbf{v}$, $\Delta\mathbf{a}$, and $\Delta\mathbf{r}$ are the relative velocity, acceleration, and position vectors of the two MBHs, respectively \citep{bellovary_first_2011}. We refer the reader to \citet{tremmel_romulus_2017} for more details on the physics modules included in the simulation.

\subsection{Galaxy Tracing}
\label{sec:galaxy_tracing}

    Halos are identified in each snapshot with the AMIGA Halo Finder spherical-overdensity algorithm \citep{knollmann_ahf_2009}. Candidate density peaks are refined adaptively; unbound particles are iteratively removed, yielding virial masses $M_{vir}$ and radii $R_{vir}$ defined where the mean enclosed density $\rho(<R_{vir})=\Delta_c(z)\rho_{crit}(z)$, where $\rho_{crit}(z)=\frac{3H^2(z)}{8\pi G}$ is the critical density of the Universe, $H(z)$ being the Hubble constant at a given redshift, and with the redshift-dependent overdensity $\Delta_{c}(z)$ computed following the top-hat collapse formalism \citep[e.g.][]{bryan_statistical_1998}. Subhalos are retained as self-bound structures within host virial radii. We calculate the centres of each galaxy/halo using the shrinking spheres approach \citep{power_2003_shrinkcentre}.

    To follow galaxies across cosmic time, galaxy merger trees are constructed by linking halos between successive snapshots based on their dark matter particles. For each halo at snapshot $s$, its descendant is identified at snapshot $s+1$ as the halo that contains the largest number of its most bound dark matter particles. This procedure yields, for each $z=0$ galaxy, a main progenitor branch and a set of secondary progenitors that merge into it. We use these merger trees to track the assembly histories of galaxies, including their stellar mass growth and merger events, and to associate MBHs with their host galaxies at all times. The stellar masses presented for MBH host galaxies are multiplied by a factor of 0.6 relative to the raw total mass of stars present in the halo. This is meant to provide a better comparison with observations, which tend to miss a substantial fraction of the true galaxy stellar mass \citep{munshi_reproducing_2013, leja_measure_2019}. This choice does not impact our results at all.

\subsection{Ex-situ and In-situ Accretion Histories of MBHs}
\label{sec:bh_trees}

MBHs are represented as collision-less sink particles in Romulus. Each MBH carries a unique identifier, its instantaneous mass ($M_{\rm BH}$), accretion rate ($\dot{M}_{\rm BH}$), host halo, position, and velocity. To quantify the relative contributions of in-situ and ex-situ processes to MBH assembly, we build MBH merger trees and decompose the mass of each black hole into components gained through gas accretion and through mergers with other MBHs.

    \begin{figure*}[]
    \begin{center}
      \centering
      \includegraphics[width=0.9\linewidth]{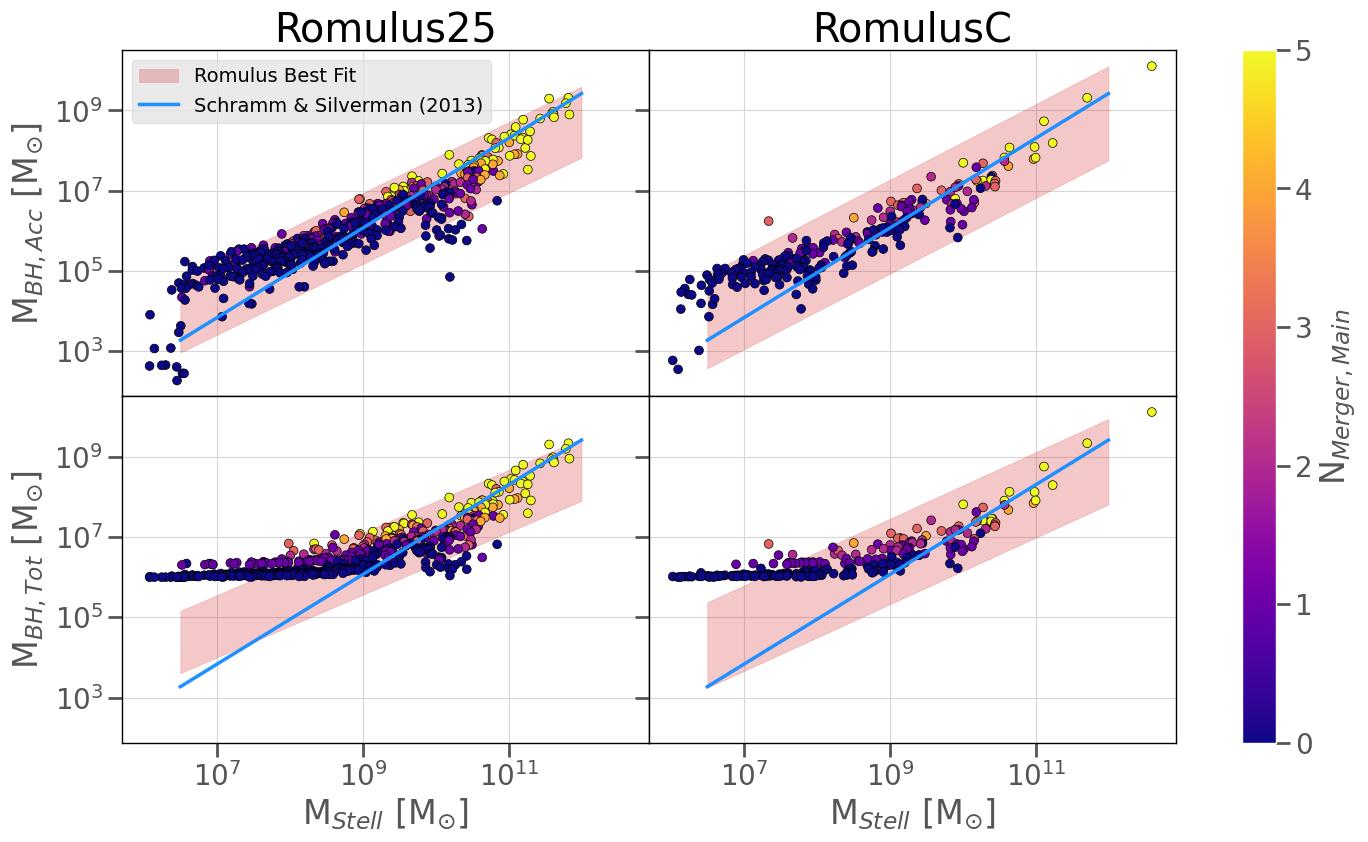} \\
    \caption{{\sc The stellar mass - black hole mass relation as a function of number of main progenitor branch mergers.} The accreted (top row) and total (bottom row) masses of central MBHs are plotted against the stellar masses of their host galaxies, with points coloured by the total number of MBH mergers occurring along the main progenitor branch. The left and right columns show Romulus25 and RomulusC, respectively. The red shaded region shows the best-fitting relation between accreted or total MBH mass and stellar mass for galaxies with $M_{\rm stell}>10^9,{\rm M_\odot}$, with the shaded region corresponding to $\pm2\sigma$ about the best fit. In the top panels, we classify central MBHs lying below the shaded region as under-grown relative to their host galaxy. The number of mergers generally increases along the relation, however, under-grown central MBHs have lower merger counts compared with the general, more massive population at a given host galaxy stellar mass. This suggests that the accretion history of a MBH is related to its merger history. The blue line shows the observed stellar mass - black hole mass relation of \citet{schramm_black_2013}.}
    \label{fig:mbh_mstar_nmerge}
    \end{center}
    \end{figure*}

We begin by linking MBH particles between consecutive snapshots using their unique IDs. A merger event is recorded when two MBH particles present in snapshot $s$ are replaced by a single MBH particle in snapshot $s+1$ that inherits one of the progenitor IDs. For each merger event, we identify the primary and secondary progenitors based on their masses at the time of merger. The full progenitor–descendant network obtained in this way defines a merger tree for each $z=0$ MBH. Each MBH's main progenitor is constructed moving backward in time from the end of the simulation. When a merger occurs, the main progenitor branch continues with the more massive black hole in that merger.

For each MBH, we have three quantities: $M_{\rm seed}$, the initial mass that is $10^6$ M$_{\odot}$ for all MBHs in the simulation; $M_{\rm acc}(z)$, the total mass gained through gas accretion up to redshift $z$, obtained by integrating the accretion rate, along an MBH's main progenitor branch, recorded in the simulation outputs; and $M_{\rm merge}(z)$, the cumulative mass contributed by MBH mergers up to $z$. Neglecting mass loss through gravitational-wave emission (which is small compared to our mass resolution), and defining $N_{\rm merge}(z)$ as the number of mergers experienced by a given MBH prior to a redshift $z$, the MBH mass at redshift $z$ can be written as
\begin{equation}
    M_{\rm BH}(z) = M_{\rm seed} + M_{\rm acc}(z) + M_{\rm merge}(z) \,.
\end{equation}
We define the ex-situ and in-situ accreted mass fractions as
\begin{equation}
    f_{\rm ex\text{-}situ}(z) \equiv \frac{M_{\rm merge}(z) - N_{\rm merge}(z)M_{\rm seed}}{M_{\rm BH}(z) - (N_{\rm merge}(z)+1)M_{\rm seed}} \,,
\end{equation}
\label{eqn: ex-situ}
\begin{equation}
    f_{\rm in\text{-}situ}(z) \equiv \frac{M_{\rm acc}(z)}{M_{\rm BH}(z) - (N_{\rm merge}(z)+1)M_{\rm seed}} \,.
\end{equation}

\noindent We also define a black hole's total accreted mass, $M_{BH,acc}$, as the total amount of mass it has gained through accretion across all of its progenitor black holes.

\begin{equation}
    M_{\rm BH,acc}(z) = M_{\rm BH}(z) - (N_{\rm merge}(z) + 1)M_{\rm seed}
\end{equation}

AMIGA associates MBHs with the halo/galaxy they are bound to. We exclude any MBH whose host galaxy is a satellite galaxy, which we define as a galaxy that is within the virial radius of another halo more than four times its virial mass. When calculating the black hole mass associated with a galaxy, we consider only the most massive black hole within $0.7~{\rm kpc}$ of the galaxy centre. Massive galaxies in particular have many wandering black holes, but these woudl in general remain undetected by conventional approaches \citep{Ricarte+2021b}. This specific distance criteria amounts to twice the gravitational softening length of the simulation. This was chosen because this is the scale below which gravity becomes unresolved and the dynamics - and the relative locations of MBHs and the potential minimum of the galaxy - become uncertain. Throughout this work, our MBH sample consists of these central MBHs, selected across all galaxies at $z=0$.
This framework allows us to follow, for each MBH and at each epoch, both its location within the galaxy and the detailed breakdown of its mass growth into in-situ gas accretion and ex-situ mergers, which forms the basis of the analysis.

\section{Results}
\label{sec:results}

    In Figure~\ref{fig:mbh_mstar_nmerge} we compare the accreted and total masses of the most massive central MBH in each galaxy with the stellar mass of its host at $z=0$, for both Romulus25 and RomulusC. The solid blue line shows the $M_{\rm BH}$–$M_\ast$ relation of \citet{schramm_black_2013}.  We present black hole mass both in terms of the total mass (bottom) and the total accreted mass, which excludes the seed masses of MBHs in the merger tree (eq. 4; top).

    To identify these under-grown systems, we fit the $M_{\rm BH, Acc}-M_{\ast}$ relation for central MBHs in galaxies with $M_{\ast}>10^9 \rm M_{\odot}$ in both Romulus25 and RomulusC. The resulting best fit relation and its $\pm2\sigma$ scatter are shown by the red shaded region in Figure~\ref{fig:mbh_mstar_nmerge}. We define an MBH as under-grown when its accreted mass lies below the lower $2\sigma$ boundary of this region, identifying MBHs that have experienced significantly less accretion relative to other MBHs in similar mass galaxies. We note that we only find such drastic under-growth in Romulus25 and not in RomulusC, though there are hits of a similar spread toward lower accreted masses at a similar mass scale, just not to the same extent. This could be due to the limited number of galaxies, or the fact that both galaxies and MBHs stop growing at earlier times due to environmental quenching \citep{tremmel_romulusc_2019,tremmel_2020,tremmel_2024_of}. 
   
    We use the accreted mass, rather than the total mass to identify these systems because previous work with Romulus has shown that the accreted component of MBH mass follows the $M_{\rm BH}$-$M_\ast$ scaling relation even below the imposed seed mass of $M_{seed}=10^6\rm M_{\odot}$ \citep{Ricarte+2019}. The relatively large seed mass imposes a floor in the total MBH mass, particularly in lower mass galaxies where MBHs may have experienced little growth. This can be seen in the bottom row of Figure~\ref{fig:mbh_mstar_nmerge}, where the total masses of the undergrown MBHs are dominated by their seed masses, making them less clearly distinguishable from the broader population. We therefore use $M_{\rm BH, Acc}$ as a more direct measure of the extent to which an MBH has grown within the simulation. For lower mass black holes in lower mass galaxies, this is also likely closer to their true mass, as $10^6$ M$_{\odot}$ is higher than theoretically expected for typical massive seed formation channels \citep[e.g.][]{volonteri_seeds_2012}.

    Despite being seeded with the same large seed mass as more typical central MBHs, the under-grown MBHs undergo significantly less accretion relative to those in similarly massive galaxies. The Romulus accretion prescription is dependent on total MBH mass, such that the large seed mass should, all else being equal, allow for similarly large accretion rates seen in the growth histories of typically massive MBHs. However, it has previously been shown that accretion onto MBHs in Romulus is most closely tied to galaxy star formation rate \citep{Ricarte+2019}, producing a similar AGN mainsequence as observed in \citet{mullaney_agnms_2012} which is nearly constant with redshift.
    If two MBHs were to remain at the centre of two different galaxies that formed the same mass in stars, one would think the MBHs should experience a similar amount of total accretion. These under-grown systems must be a result of unique conditions causing them to deviate from this relationship.

\subsection{In-Situ vs. Ex-Situ Growth}
\label{sec:exsitu}
    \begin{figure}[h]
    \begin{center}
      \centering
      \includegraphics[width=.9\linewidth]{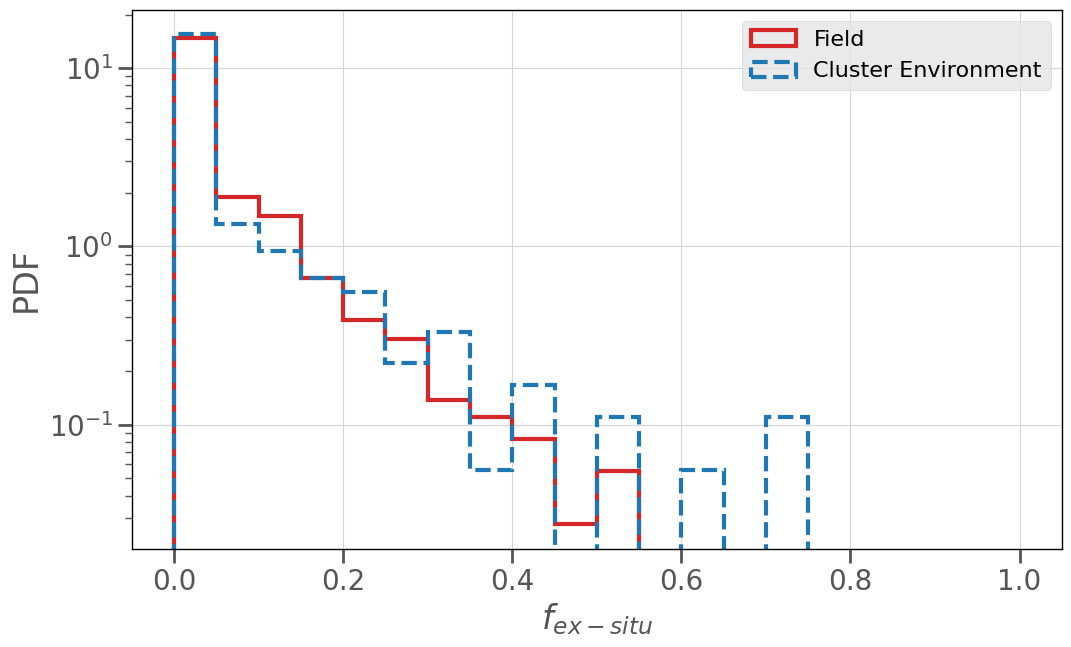} \\
    \caption{{\sc Distribution of ex-situ accretion among MBHs.} The normalised distributions of the fraction of a black hole's accreted mass which has been accreted ex-situ for the most massive, central black holes in galaxies for the field (red; Romulus25) and cluster (blue; RomulusC) environments. The distributions are highly similar, with both displaying peaks at zero and quickly dropping for higher fractions.}
    \label{fig:exsitu_env}
    \end{center}
    \end{figure}

    \begin{figure}[]
    \begin{center}
      \centering
      \includegraphics[width=.9\linewidth]{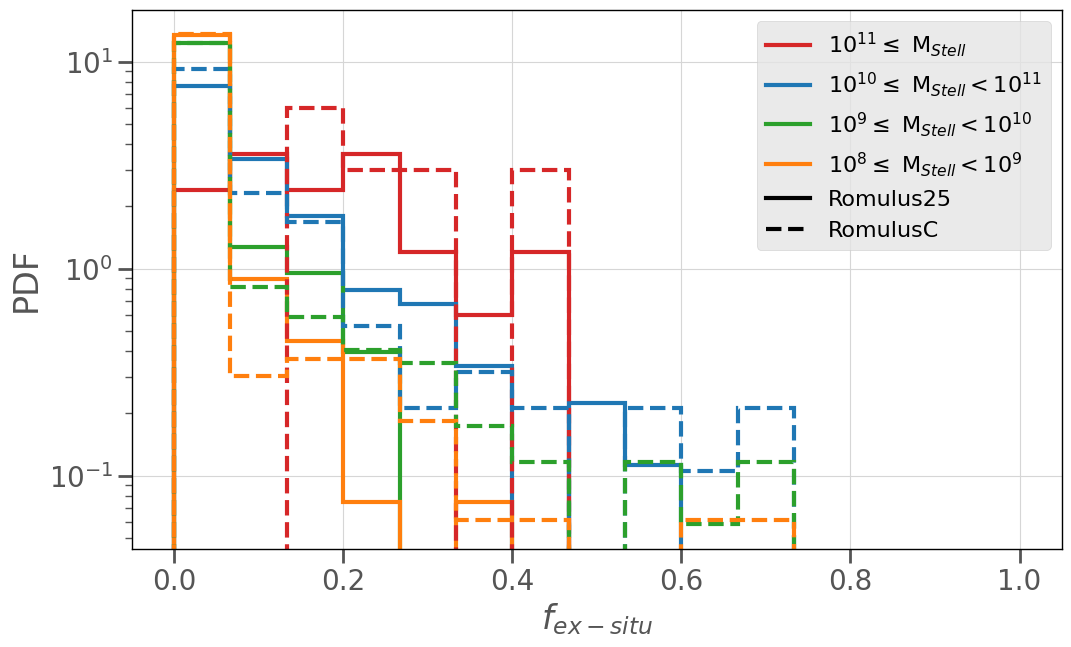} \\
    \caption{{\sc Comparison of the ex-situ growth for MBHs for different stellar mass bins in the field and cluster populations.} The normalised distributions of the fraction of a black hole's total accretion which is ex-situ for the most massive central black holes in galaxies in the field (solid; Romulus25) and cluster (dashed; RomulusC) environments, divided into four host galaxy stellar mass bins. The distributions are similar in both environments, with the three lower stellar mass bins displaying peaks at zero and the highest stellar mass bin peaking at a higher fraction of approximately 0.2. higher stellar mass bins also display wider distributions than the lower mass bins. MBHs typically gain most of their mass from in-situ accretion regardless of mass or environment.}
    \label{fig:exsitu_mass}
    \end{center}
    \end{figure}

    One obvious difference between the under-grown population and that of their peers in similar-mass galaxies is the number of mergers they experience. With this in mind, we examine how important ex-situ accretion is to MBHs in galaxies across environment and galaxy mass. That is, how much of the total accreted mass is contributed by accretion events taking place in MBHs that have merged with the primary MBH in the past. Figure~\ref{fig:exsitu_env} shows the normalised distributions of the ex-situ growth fraction, defined in equation ~\ref{eqn: ex-situ}, for the most massive central MBH in each galaxy at $z=0$. We compare systems drawn from the uniform-volume Romulus25 run, which we treat as a field environment, to those in the RomulusC cluster zoom, which can be considered a high density environment. We note that our comparison with a cluster environment is based on the single RomulusC cluster zoom-in. As a result, the trends identified for cluster galaxies may depend on the particular assembly history of this system, and should not be taken as representative of the full cluster population. However, comparing the two is still instructive. In both cases the distributions are sharply peaked at low ex-situ contributions, with the majority of MBHs experiencing negligible accreted mass coming in through mergers with only a small tail having ex-situ fractions greater than $0.1$. We performed a two-sample Kolmogorov-Smirnov test between the two populations, obtaining a p-value of 0.454, indicating no statistically significant evidence that the two distributions differ. Galaxies in the cluster simulation tend to lose their gas via ram pressure stripping, quenching both star formation and MBH accretion \citep{tremmel_romulusc_2019, Ricarte+2020, tremmel_2020}. Similarly, mergers also become less common in cluster environments, as galaxies move too quickly relative to one another to merge and bring MBH pairs together. So, both galaxies and MBHs experience assembly bias, whereby objects of a given mass (MBH mass, stellar mass) must have grown to that mass earlier (i.e. before in-fall) in the cluster environment compared to the field \citep[e.g.][]{boylankolchin_2009}. These results tell us that, whether a MBH assembles its mass early (cluster) or later (field), it is likely to have little ex-situ accretion.
    
    We next examine how the balance between in-situ and ex-situ MBH growth depends on host galaxy stellar mass. Figure~\ref{fig:exsitu_mass} shows the normalised distributions of the ex-situ accreted mass fraction, split into four bins of host stellar mass from $10^8$ to $\gtrsim 10^{11}\,{\rm M}_\odot$, for the most massive central MBH in each galaxy in Romulus25 (solid lines) and RomulusC (dashed lines). In both environments the three lowest-mass bins ($10^8 \leq M_\ast < 10^{11}\,{\rm M}_\odot$) are strongly peaked at zero ex-situ contribution, indicating that MBHs in low- and intermediate-mass galaxies grow almost entirely through in-situ gas accretion. The distributions in these bins decline steeply beyond ex-situ fractions of a few per cent, with only a small number of systems reaching $f_{\rm ex\text{-}situ} \gtrsim 0.1$.

    In contrast, MBHs hosted by the most massive galaxies ($M_\ast \geq 10^{11}\,{\rm M}_\odot$) show broader distributions that are shifted toward higher ex-situ fractions. In both Romulus25 and RomulusC the peak for this bin occurs at $f_{\rm ex\text{-}situ} \sim 0.2$ with some systems as high as $\sim0.4$. This trend reflects the increasing importance of black hole mergers in the assembly of the most massive systems. The close similarity of the mass-binned distributions between Romulus25 and RomulusC further suggests that this dependence on host stellar mass is independent of whether the MBH resides in a field or cluster environment. However, our under-grown population defined in the previous section fits into the $10^{10-11}$ M$_{\odot}$ bin, which is dominated by low amounts of ex-situ accretion. Even the highest mass bins show typical fractions below 0.3. This can therefore not explain the significant difference in the under-grown population. This means that the lack of mergers is not the direct cause, but potentially another symptom related to this lack of growth.
    
\subsection{Undermassive Black Holes as a Consequence of Complex Dynamical Histories}
\label{sec:undermassive}

    \begin{figure*}[]
    \begin{center}
      \centering
      \includegraphics[width=.9\linewidth]{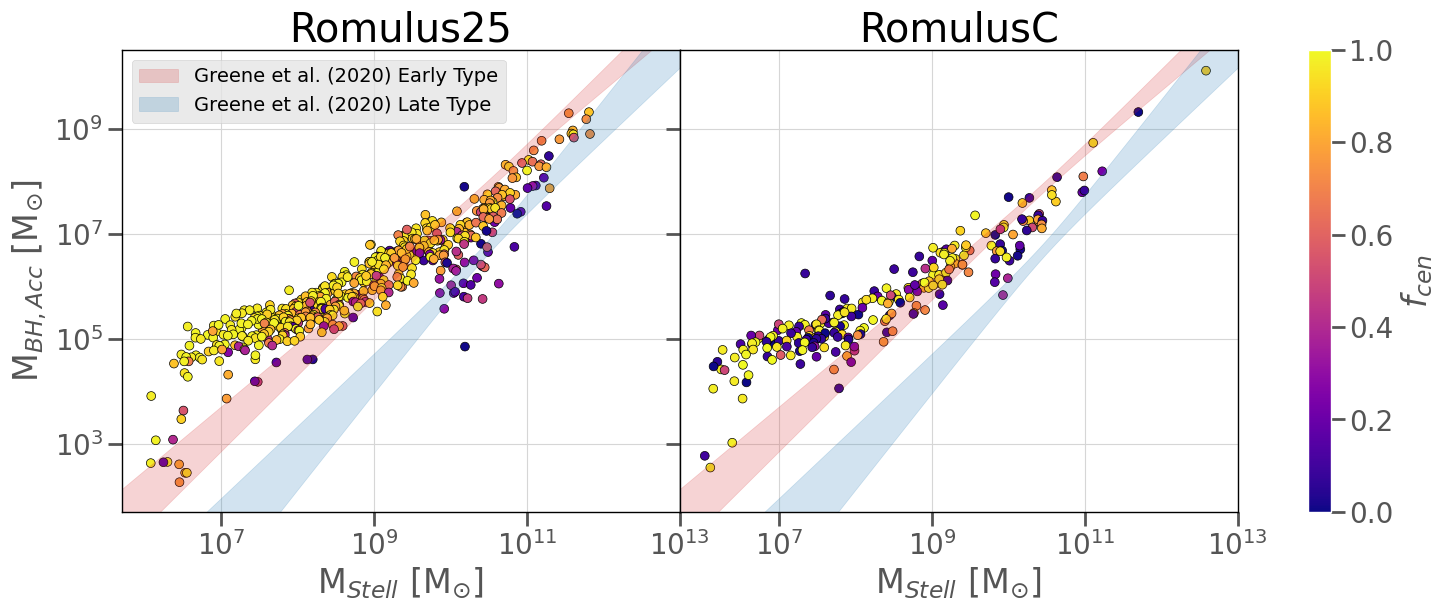} \\
    \caption{{\sc The stellar mass - black hole mass relations and the fraction of time a MBH spends at galaxy centre.} The accreted mass of central MBHs are plotted against the stellar mass of their host galaxies, with data points coloured by the fraction of its lifetime that a black hole has spent in the central 700 pc of its current host galaxy, $f_{\rm cen}$. The red and blue shaded regions show the $M_{\rm BH}-M_{\rm stell}$ scaling relations for early- and late- type galaxies, respectively, from \citet{greene_2020}. For the Romulus25 population, $f_{\rm cen}$ is typically high, above approximately 70\% for data points on or near the relation. However, for under-grown MBHs below the relation, the fraction is lower than their counterparts in similar mass galaxies. For the RomulusC population, a similar but much weaker trend is seen.}
    \label{fig:mbh_mstar_tcfh}
    \end{center}
    \end{figure*}

    \begin{figure*}[]
    \begin{center}
      \centering
      \includegraphics[width=.99\linewidth]{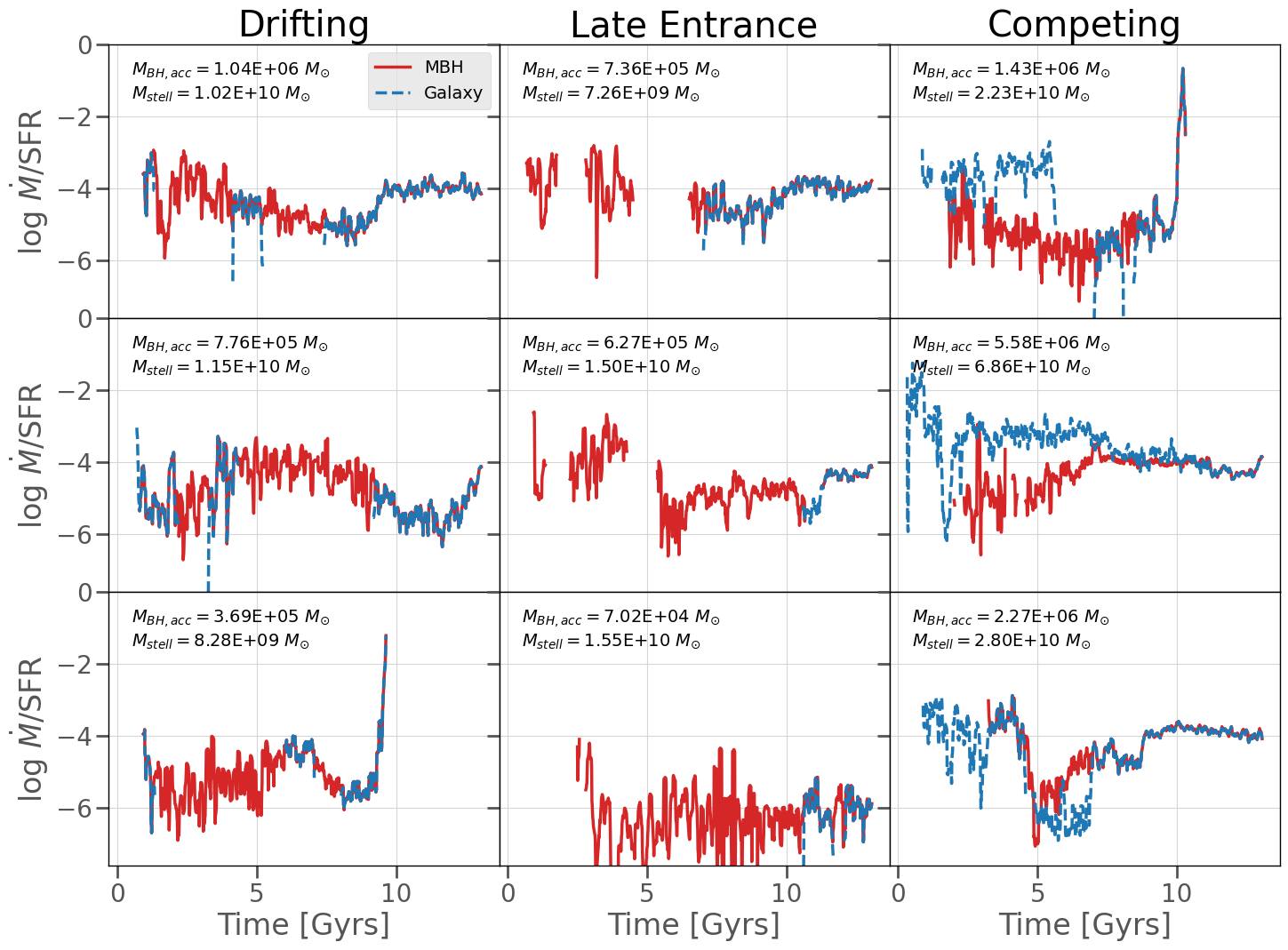} \\
    \caption{{\sc Examples of three archetypical dynamical evolution scenarios leading to under-grown MBHs.} The ratio of a black hole's accretion rate to host galaxy star formation rate versus time, traced along the black hole's main progenitor branch (red) and along the final host galaxy's main progenitor branch (blue). Each panel is labelled with final MBH accreted mass and host stellar mass. Each MBH is under-grown compared to the general population at the same galaxy mass. Each of the under-grown MBHs in the field environment display one of the types of evolutions shown. For the top right and bottom left panels, no data is shown for the last \textasciitilde 2 Gyr, and for the centre top and middle panels there are gaps seen at \textasciitilde 5 Gyr, in each case this is due to the galaxy SFR being 0. In all scenarios shown, the MBH spends the majority of its lifetime accreting at substantially lower values than what would be required to exist on typical scaling relations.}
    \label{fig:sfr_bhar_evolution}
    \end{center}
    \end{figure*}

    \begin{figure*}[]
    \begin{center}
      \centering
      \includegraphics[width=.9\linewidth]{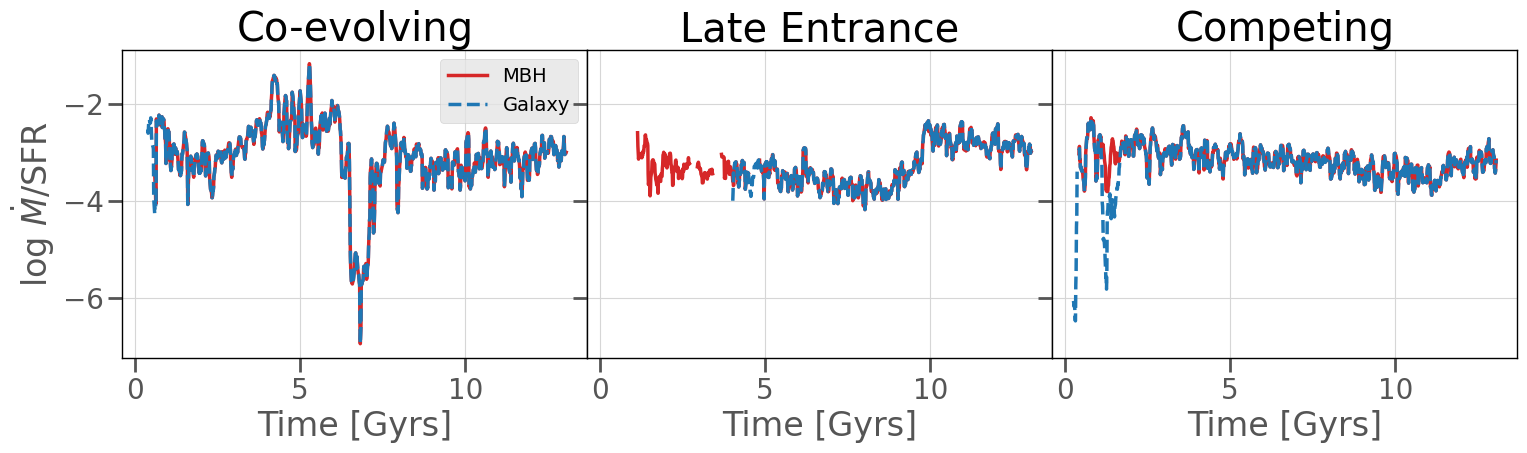} \\
    \caption{{\sc Examples of growth histories for MBHs that have typical masses.} The ratio of MBH accretion rate to host galaxy SFR over time, tracking along the MBH's main progenitor branch (red) and the galaxy's main progenitor branch (blue). Mass matched examples from Romulus25 to the galaxies presented in Fig.~\ref{fig:sfr_bhar_evolution} selected to have more typical final accreted masses. While co-evolutiom is the most common evolutionary path (left), we find examples of `late entrance' and `competing' MBHs, but these phases end much earlier than the under-grown population.}
    \label{fig:sfr_bhar_evolution_control}
    \end{center}
    \end{figure*}

    \begin{figure}[h]
    \begin{center}
      \centering
      \includegraphics[width=.9\linewidth]{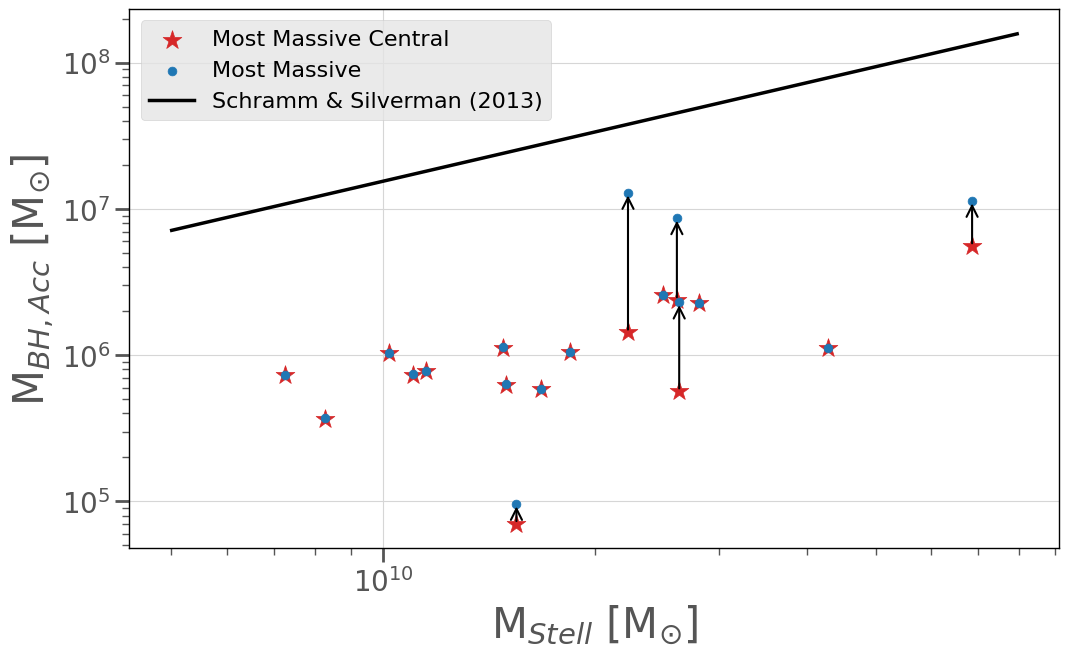} \\
    \caption{{\sc Some galaxies with under-grown MBHs have more massive wandering MBHs in them.} The relation between the total MBH accreted mass for central, under-grown MBHs and their host galaxy stellar mass is shown in red stars. Arrows are drawn between them and the most massive, wandering MBHs (blue circle) in the same galaxy. Only two galaxies in this sample have an off-centre MBH with an accreted mass substantially closer to the empirical relation.}
    \label{fig:mmc_vs_mm}
    \end{center}
    \end{figure}

    \begin{figure*}[h]
    \begin{center}
      \centering
      \includegraphics[width=.99\linewidth]{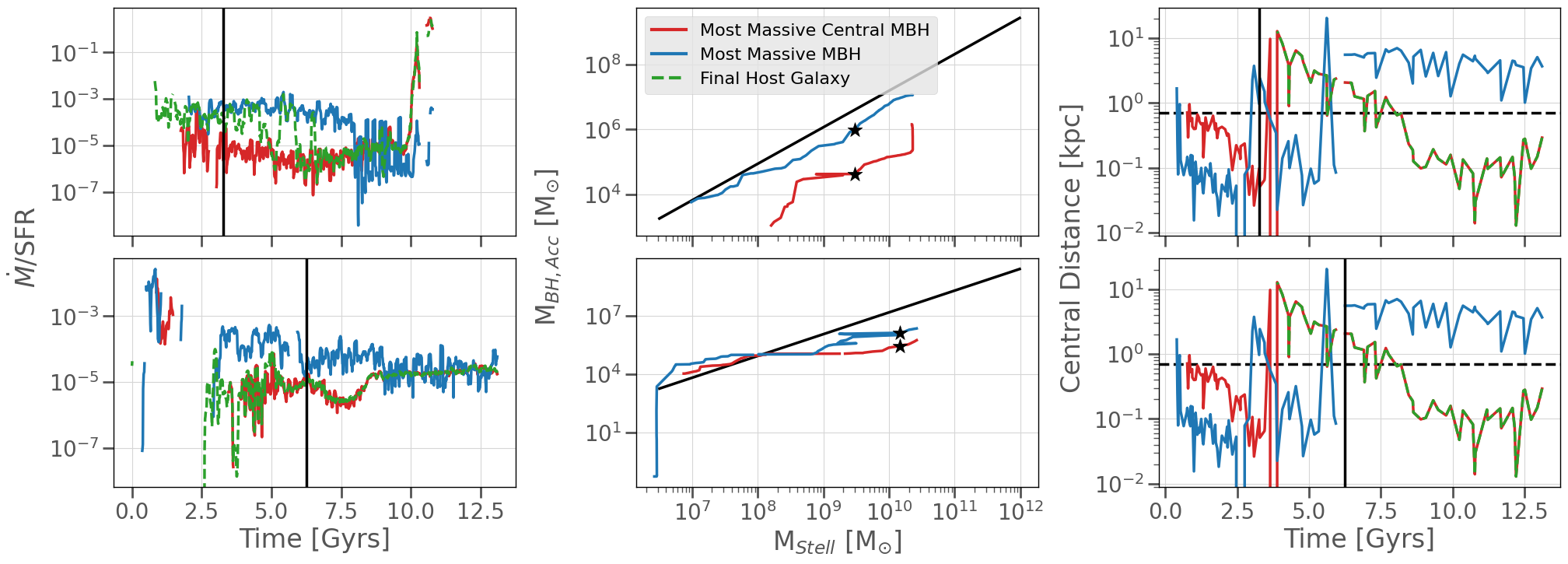} \\
    \caption{{\sc Two examples of galaxies that have their most massive MBHs off-centre.} \textit{Top:} a galaxy with a wanderer that has a typical accreted mass for its galaxy's stellar mass. \textit{Bottom:} an example where the off-centre MBH is still substantially under-grown by $z=0$. In each panel we follow the MBH that is the most massive, central MBH at $z=0$ (red), the MBH that is the most massive in the entire galaxy at $z=0$ (blue), and the most central MBH along the galaxy's main progenitor branch (green, dashed). The left column shows the ratio of the MBH accretion rate to the host galaxy SFR as a function of time. The middle column shows the evolution of accreted MBH mass with host galaxy stellar mass. The right column shows the distance of each MBH from the centre of its host galaxy as a function of time. The galaxy merger that brings the two $z=0$ MBHs into the same host is indicated by a vertical black line in the left and right panels at the time of the merger, and by black stars along their evolutionary tracks in the middle panels. Apparent sudden decreases in stellar mass in the middle panels occur during galaxy mergers, when the halo finder can briefly reassign particles between the merging halos.}
    \label{fig:offnuclear}
    \end{center}
    \end{figure*}


    We next explore whether the under-grown central MBHs are associated with atypical orbital histories within their present-day hosts. For each MBH we compute the fraction of its lifetime that the main progenitor MBH has spent within the central $700~{\rm pc}$ of its current host galaxy, $f_{\rm cen}$. Since the MBH position within its host is only available at discrete simulation outputs, we approximate $f_{\rm cen}$ by assuming that an MBH's classification (central or non-central) at any snapshot has been true since the previous snapshot. Figure~\ref{fig:mbh_mstar_tcfh} shows the $M_{\rm BH}$–$M_\ast$ plane for Romulus25 (left) and RomulusC (right), colour-coded by $f_{\rm cen}$. The early- and late-type galaxy scaling relations from \citet{greene_2020} are shown for comparison. In Romulus25, MBHs that lie on or close to the early-type relation typically have high central fractions, $f_{\rm cen} \gtrsim 0.7$, indicating that they have spent most of their lifetimes near the centres of their present-day hosts. Along the relation, $f_{\rm cen}$ decreases only mildly toward higher masses, suggesting that massive, well-grown MBHs still spend a substantial fraction of their history in the nuclear region.

    By contrast, the under-grown central MBH population extend away from the early-type relation and towards the late-type relation of \citet{greene_2020}. These systems  have lower central fractions, $f_{\rm cen} \lesssim 0.6$, indicating more extended periods of wandering away from galactic centre, which may have limited their ability to accrete gas. This suggests a possible connection between the position of the MBH relative to observed scaling relations and its orbital history: prolonged off-centre evolution may reduce access to the dense nuclear gas required for efficient growth, resulting in a shift towards the lower normalisation of the late-type relation. We stress, however, that proximity to the early- or late-type relations does not itself imply a corresponding morphology for the simulated host galaxies. In RomulusC a similar, trend is present, but as discussed previously we lack the more substantially under-grown population seen in Romulus25.
    

    Looking in more detail at the orbital histories of our sample of significantly under-grown MBHs (more than 2-$\sigma$ below the best-fit relation) and find three archetypical examples that result in under-grown central ($D<0.7$ kpc) MBHs by $z=0$. To aid in this classification we differentiate between tracking a MBH along is main progenitor branch and its host galaxy at $z=0$. Figure~\ref{fig:sfr_bhar_evolution} shows how the ratio of black hole accretion rate to star formation rate (SFR) over time for three examples within each archetypical class of under-grown black holes. The red lines show the evolution of the most massive central MBH at $z=0$ along its main progenitor branch, with the star formation rate taken from its current host at any given time. The blue dashed line tracks the galaxy's main progenitor and its most massive, central MBH at any given time. Periods without a blue line indicate times when the galaxy has no central MBH. Times when the lines do not lie on top of one another indicate periods where a different MBH is in the centre of the tracked galaxy.

    The left column of Figure~\ref{fig:sfr_bhar_evolution} shows \emph{drifting black holes}. Here, the traced MBH is central in the final host at multiple times, but its orbit carries it repeatedly in and out of the central region. When the MBH is off-centre, the galaxy trace indicates no central black hole (marked by periods without a blue dashed line), while the BH trace shows the MBH, in most cases still in the main galaxy progenitor, growing away from the centre at a rate that is far below the typical ratio given by the AGN Main Sequence of $\sim10^{-3}$ \citep{mullaney_agnms_2012}. In some cases, such as that shown in the bottom left panel, the MBH becomes central and experiences rapid growth, eventually extinguishing star formation within its host galaxy (hence why the lines stop at around 10 Gyr).
    
    The centre column shows \emph{late-entrance black holes}. In this case, the MBH joins the final host only at relatively late times. For roughly the first half of the galaxy's evolution there is no MBH within 0.7 kpc of the centre. The MBH that will eventually become central is usually in a different galaxy entirely, falling into the primary halo later and only arriving at the centre after many Gyr of evolution. While the MBH is off centre, it has a difficult time accreting effectively. Once the MBH arrives at the centre, its accretion history is too short to bring it up to the mass expected for a central MBH of the host galaxy, leaving it under-grown at $z=0$.
    
    The right column shows an example of systems with \emph{competing black holes}. Here the traced MBH is central in the final host for a substantial fraction of its lifetime, but another MBH is also present and alternately becomes the most massive central black hole. Whichever MBH is at the centre is generally able to grow more. In this sense, the black holes are competing for gas resources and neither is able to grow effectively. In the upper right panel we see an example of a competing MBH that becomes more stable and then grows rapidly, quenching its host galaxy.
    
    Across the under-grown central MBH sample in Romulus25, all systems exhibit one of the three evolutionary patterns described above. In every case, the final host galaxy experiences significant periods with either no central MBH or a central black hole different from the one we identify at $z=0$, allowing substantial stellar mass growth with limited MBH growth.


    For comparison, we repeat the same analysis for a control sample of ``typically grown'' central MBHs in Romulus25, selected to have similar host stellar masses to the under-grown systems but to lie closer to the general population on the $M_{\rm BH,acc}$–$M_\ast$ relation. Figure~\ref{fig:sfr_bhar_evolution_control} shows representative examples of the evolution of $\log(\dot{M}_{\rm BH}/{\rm SFR})$ for this sample, again contrasting the BH trace with the galaxy trace. The most common behaviour, illustrated in the left panel, is that of a \emph{co-evolving black hole and galaxy}: the MBH is central in the final host for essentially its entire lifetime. In these systems the growth of the black hole tracks the star-formation history of the host with no extended episodes in which the galaxy lacks a central MBH or has a different MBH in the centre. The middle panel shows a rarer \emph{late-entrance} case, in which the MBH joins the final host slightly later. However, unlike the under-grown systems, the duration of this phase is relatively short, with the MBH arriving at the centre early on and, from there, it co-evolves with the host. Finally, the right panel presents a \emph{competing black hole} example, where a second MBH is present and briefly rivals what will become the largest central MBH. However, unlike the under-grown population, this competition phase is shorter lived and is resolved (i.e. the primary MBH becomes central and remains there) at much earlier times.
    
    Across the ``typically grown'' sample, essentially all systems fall into one of these three categories, with the co-evolving class by far the most common. In contrast to the under-grown central MBHs, long intervals during which the final host has no central black hole, or during which the eventual central MBH is off-centre or sub-dominant, are rare. Also, the ratio of MBH accretion and star formation is more consistently $\sim10^{-3}$, following roughly the expected ratio to produce the $z=0$ empirical relation. But this co-evolution can only occur if the MBH is dynamically stable at the centre of its host galaxy.


    To assess whether the under-grown classification is driven by our requirement that central MBH be central, we compare, for each under-grown galaxy, the most massive \emph{central} MBH to the most massive MBH \emph{anywhere} in the galaxy. Figure~\ref{fig:mmc_vs_mm} shows $M_{\rm acc}$ versus $M_\ast$ for the under-grown sample, with red stars indicating the most massive central MBH and blue circles the most massive MBH in the entire galaxy. For the majority of systems the two symbols coincide, demonstrating that the most massive black hole is also central and that the under-grown nature of these central MBH is not an artefact of neglecting an off-centre MBH that had a richer accretion history.

    An exception occurs in four hosts in the sample, which each contain an off-centre MBH that has grown more than any central black hole. In two of these galaxies the off-centre MBH would still be considered under-grown. In the other two, the off-centre MBH is significantly more massive and lies much closer to the empirical relation. This case illustrates that, in a small subset of galaxies, a wandering or displaced MBH could in principle reconcile the host with the observed $M_{\rm BH}$–$M_\ast$ relation, but for the majority of our under-grown systems no such hidden massive black hole exists.


    We now examine in more detail the two systems in which the most massive MBH is off-centre at $z=0$. Figure~\ref{fig:offnuclear} compares the evolution of the most massive central MBH at $z=0$ (MMC), the most massive MBH in the galaxy at $z=0$ (MM), and the MBH closest to the galaxy centre at each simulation output (galaxy trace). Importantly, the MM and MMC do not originate in the same host. The vertical black lines in the left and right columns, and the black stars in the middle column, mark the galaxy merger that brings the two MBHs into the same system.

    The top row shows a system in which the off-nuclear MM reaches a mass broadly consistent with that expected for the final host, while the MMC remains substantially under-grown. Prior to the galaxy merger, the two MBHs evolve independently within their respective hosts. At the time of the merger, the MM has already grown substantially more than the MMC, as indicated by their different accreted masses in the middle panel. Following the merger, the MM remains the more massive of the two but is displaced from the nucleus, spending much of its subsequent evolution several kiloparsecs from the galaxy centre. The galaxy trace therefore switches between both MBHs, and for a brief time, a third smaller MBH, depending on which is most central. At late times, the MMC settles within the central 0.7 kpc and its accretion rate relative to the host SFR increases, however, this enhanced growth occurs too late for it to reach the mass of the MM by $z=0$.

    The bottom row provides a contrasting case in which both the MM and MMC remain under-grown relative to the final host. Here the galaxy merger bringing the two MBHs together occurs later in the systems evolution, after the host has already undergone substantial stellar mass growth. By the time of the merger, the host stellar mass is already close to its present day stellar mass, and both MBHs  lie well below the observed $M_{\rm BH}$–$M_\ast$ relation. The subsequent merger of their host galaxies does not lead to sufficient sustained accretion for either MBH to reconcile their mass with the relation. The MM remains several kpc from the centre for much of the post-merger evolution, while the MMC does sink to the centre but continues to grow weakly.

    While galaxy mergers also increase the stellar mass of a system without necessarily producing a corresponding increase in MBH mass, both systems shown in Figure~\ref{fig:offnuclear} are already under-grown relative to their hosts prior to the merger. This indicates that the merger itself is not the primary origin of their under-grown masses.

    These two systems show how the present-day configuration can retain information about the assembly histories of both the MBHs and their hosts. In the first case, the galaxy merger introduces an MBH that had already grown efficiently in its original host, leaving the remnant galaxy with a typically massive but off-nuclear MBH alongside an undermassive central MBH. In the second case, both MBHs have already fallen behind the growth of the final host when they are brought into the same galaxy, and both remain undermassive thereafter. Although the addition of stellar mass through galaxy mergers, without corresponding MBH growth, may contribute to the offset from the $M_{\rm BH}$–$M_\ast$ relation, these examples demonstrate that the undergrown nature of these central MBHs cannot be attributed solely to the assembly of their present-day host. Instead, their under-grown masses were already established in their progenitor galaxies. Whether an off-nuclear MBH can account for the apparent mass deficit of the present day central MBH depend in part on the growth histories of the separate progenitor systems and on when they are assembled into the same galaxy.


\section{Discussion}
\label{sec:discussion}

\subsection{Connection to the wandering black hole population}

The dynamical decoupling between galaxy and MBH growth histories that we identify is closely related to the
population of \emph{wandering} black holes previously characterized in the Romulus simulations by \citet{Ricarte+2021a}. They used the same simulation suite (and the same $0.7$\,kpc central-aperture criterion adopted in our $f_{\rm cen}$ analysis) to demonstrate that a substantial fraction of MBHs in Romulus25 and RomulusC spend significant fractions of cosmic time offset from their host galaxy
centres. \citet{tremmel_2018_wanderers} similarly demonstrated that wandering is common even in Milky Way-mass halos. Our results extend this picture by showing explicitly \emph{how} such wandering translates into an observable deficit relative to the $M_{\rm BH}$--$M_*$ relation: it is not merely that wandering black holes exist, but that the specific dynamical history of the black hole that happens to be central \emph{today} determines whether a central MBH appears under-grown.

Observational results such as \citet{greene_2020} and \citet{reines_relations_2015} find populations of MBHs that are below typical scaling relations, identifying that multiple scaling relations might be at play. The under-grown black holes from \citet{greene_2020} are in galaxies of similar stellar mass as the ones hosting the under-grown MBHs in Romulus. It is possible that dynamical histories like those presented here could be the cause for an increased scatter in MBH masses. This mass scale of $\sim10^{10-11}$ might represent a middle ground between galaxies hosted by massive enough dark matter halos to experience rich merger histories, keeping or perturbing MBHs away from centre, but not so massive as to prevent such perturbations from happening in the first place due to a deeper potential well. In the \citet{reines_relations_2015} sample, under-massive black holes in these galaxies are more likely to shine as AGN. We see in multiple of our examples that once the under-grown MBH reaches the centre it can begin to grow very rapidly, as the galaxy has been able to accumulate a lot of gas without feeling the impact of AGN feedback. It is still possible that multiple scaling relations exist and we stress that Romulus was optimized to reproduce a single scaling relation with stellar mass. However, it is important to consider that a diversity of dynamical histories can impact the scatter in these relations.

This work also connects directly to a rapidly growing observational literature on wandering and off-nuclear massive black holes. \citet{Mezcua+2020} reported signatures of off-nuclear AGN activity among MaNGA dwarf galaxies. MBH pairs have been detected at sub-kiloparsec separations \citep{voggel_2022}, adding to the various evolutionary phases captured in this galaxy-scale dynamical evolution. \citet{Ricarte+2021b} predicted that a subset of the wandering population in simulations such as Romulus should be directly detectable via tidal disruption events (TDEs) offset from their host galaxy nuclei; this prediction has since been borne out observationally, with \citet{Guolo+2026} presenting a detailed spectral energy distribution analysis of the off-nuclear TDE AT2024tvd that confirms a wandering supermassive black hole with no associated stellar overdensity. Further, \citet{Guolo+2026b} show that the host galaxies and off-nuclear distances of several TDEs are consistent with predictions from the Romulus simulations presented in \citet{Ricarte+2021b}. While observations generally focus on finding wandering MBHs at low redshift, JWST is beginning to detect galaxies with off-nuclear AGN \citep{ubler_offcen_2024} as well as galaxies with multiple MBHs \citep{ubler_tripleBH_2026} at early times. Our results show that the dynamical histories of MBHs at early times can significantly impact their final accretion histories and masses in the local Universe. An initially off-centre black hole might remain undergrown (drifting), or a multi-MBH system identified at early times could remain so, resulting eventually in an under-grown central MBH once the dynamics settle down (competing).

\subsection{Comparison to other cosmological simulations}

The magnitude of ex-situ growth we find for the most massive hosts is qualitatively, though not quantitatively, consistent with results from other large-volume cosmological simulations. Using IllustrisTNG, \citet{weinberger_simulating_2017} found that black holes above $M_{\rm BH} \sim 10^{8.5}\,M_\odot$ grow \emph{most} of their mass through mergers once they transition to the low-accretion-rate kinetic feedback mode, a substantially stronger merger dependence than the $f_{\rm ex\text{-}situ} \sim 0.2$ peak we find for the highest-mass bin in Romulus25. This quantitative difference likely reflects differences in sub-grid AGN feedback prescription - IllustrisTNG's kinetic-mode feedback strongly suppresses in-situ gas accretion once the accretion rate falls below a threshold. This suggests that the relative balance of
in-situ and ex-situ growth predicted for the most massive black holes is sensitive to feedback-model choices in a way that is not yet fully disentangled from the treatment of BH dynamics and merging. However, it should also be noted that, in addition to feedback and accretion, IllustrisTNG employs simple prescriptions for MBH seeding and dynamics. MBHs fail to form early in the simulation due to this seeding algorithm while at later times MBH mergers occur unrealistically fast due to artificial repositioning that fails to account for MBH dynamical evolution.

The Astrid simulations \citep{ni_astrid_2022} employ a similar dynamical friction model as Romulus \citep{chen_df_2022}, finding wandering MBH populations even at very early times \citep{dimatteo_2023_wanderers}. Like in Romulus, these wandering MBHs experience less accretion than their central counterparts, so it would stand to reason that Astrid might also have a similar population of under-grown MBHs. Unlike Romulus, their MBH seed masses are much smaller and realistic, though the impact of dynamical friction is artificially boosted at the lowest masses. The New Horizon simulations have been shown to have very inefficient black hole growth in low mass galaxies due to MBHs becoming perturbed away from galactic centre, similar to our `drifting' population presented here \citep{beckmann_2023}.

It has been previously established in zoom-in and idealized simulations that the dynamics of black holes following galaxy mergers are complex. For example, \citet{vanWassenhove_2014} find that black hole pairs can have complex dynamical histories as a result of galaxy mergers, finding that it is possible for the black hole in a lower mass, secondary galaxy to replace the black hole and stellar nucleus of the primary galaxy. This effect is similar to what is seen here for competing MBHs.

\subsection{Connection to Quenching Star Formation}

In a couple of our case studies (see Figure~\ref{fig:sfr_bhar_evolution}) we find that black holes undergo a period of rapid growth relative to their galaxies, eventually quenching them all together. This can be due to the fact that the otherwise underactive black hole results in a galaxy that is very gas-rich in the centre. Once the black hole settles in the centre, it is able to grow more rapidly. These two examples indicate that a sub-set of these under-grown MBHs that could be detected as AGN could also be in the process of quenching their host galaxies. In lower mass galaxies within the Romulus25 simulation, \citet{sharma_2023} find that under-grown black holes are less likely to impact their host galaxy. Our results do not contradict this work, but show that it is possible that late-time accretion onto an otherwise under-grown black hole can have an impact in higher mass galaxies. It makes sense for this impact to be mass-dependent. The gas supply in lower mass galaxies is more impacted by star formation and supernovae feedback, which also limites MBH growth \citep{dubois_2015,habouzit_2017}. At higher masses, supernovae feedback is less efficient and AGN feedback becomes more important. Thus, these galaxies with under-grown MBHs are likely to accumulate more gas mass that is then made available for accretion once a MBH finds its way to the centre. \citet{sanchez_2024} found that MBHs above the $M-\sigma$ relation were more efficient at driving outflows in Romulus galaxies, as indicated by the enriched circumgalactic medium. While we do not attempt here to examine the under-grown MBHs with respect to the $M-\sigma$ relation here, nor characterize their outflows, the examples of under-grown MBHs that seem to cause quenching are few and would not contribute significantly to the overall population. However, it could still be interesting to investigate these specific, under-grown systems in this context in future work.

\subsection{Limitations and future directions}

Throughout this work we have relied on the accreted mass of MBHs. However, because accretion follows a Bondi-Hoyle-like routine, it is highly dependent on the mass of the black holes, which are seeded at $10^6$ M$_{\odot}$. At the lowest values for M$_{\rm BH,acc}$, the actual mass of the black hole will also be highly dependent on this initial mass. While the details of MBH seed formation are uncertain, leading models generally predict an upper limit of $10^5$ M$_{\odot}$ for massive seed formation scenarios \citep{volonteri_seeds_2012}. It is likely that the earliest phases of MBH growth could be overestimated in Romulus, given the large seed masses. However, the limited resolution of Romulus also results in an extended period of MBH formation, ending at $z\sim5$, much later than typically thought for MBH seed formation. Still, the specific values of $M_{\rm BH, acc}$ at the lowest galaxy masses predicted from Romulus should not be treated with great certainty. Rather, what is interesting for this work is the comparative analysis of MBH accretion histories within galaxies of a given mass/environment and how they relate to the overall relationship between M$_{\rm BH,acc}$ and M$_{\star}$ predicted from the simulation and not the specific values.

Looking ahead, a natural extension of this work is to connect the dynamical assembly histories identified here to the black hole spin demographics (Whitaker et al., in prep.). Because spin evolution is most efficient during sustained, coherent nuclear accretion, the drifting, late-entrance, and competing black hole pathways identified in Section~4, all of which involve extended periods of reduced or displaced central accretion, offer a directly testable prediction: under-grown AGN should preferentially retain lower spins than typically massive systems on the $M_{\rm BH}$--$M_*$ relation, since they spend less of their growth history embedded in a coherent nuclear gas reservoir. Testing this prediction with spin calculated in post-processing would determine if observed MBH spin measurements could be used to distinguish MBHs with complex dynamical histories like those shown here. This would build on other work using the Horizon and NewHorizon simulations which have shown how MBH spin can encode the accretion and merger histories of MBHs \citep{beckmann_2024,beckmann_2025}.

\section{Conclusions}
\label{sec:conclusions}

Using the Romulus25 and RomulusC cosmological simulations, we have examined the importance of mergers and dynamics to the growth of MBHs in the Romulus simulations. We found:

\begin{itemize}

\item We identified 17 under-grown central MBH identified in Romulus25 compared with other central MBHs of similar mass galaxies. These MBHs were characterized by both a quiet merger history and spending significant time away from galactic centre during their history.

\item Ex-situ growth is a minor contributor to MBH assembly $M_* \lesssim 10^{11}\,M_\odot$, becoming appreciable only for the most massive hosts ($f_{\rm ex\text{-}situ} \sim 0.2$ for
$M_* \gtrsim 10^{11}\,M_\odot$), with no dependence on environment (field versus cluster). Even in massive galaxies with MBHs that experience richer merger histories, MBHs experience most of their accretion along their main progenitor branch. The lack of mergers cannot explain the under-grown population seen in the simulation.

\item Tracing the growth histories of under-grown MBHs and their host galaxies reveals three distinct evolutionary pathways, through which a galaxy's final central MBH growth history can be decoupled from the bulk of its host's star formation history: drifting, late entrance, and competing black holes. MBHs in such systems experience significantly lower accretion relative to star formation due to the limited gas supply away from galaxy centre. MBHs which spend more time at the centres of their galaxies experience growth that is reminiscent to the AGN Main Sequence ($\dot{M}_{\rm BH}/SFR \sim 10^{-3}$, leading to MBHs that lie close to the empirical relations. 

\item We do not find such significantly under-grown MBHs in the RomulusC simulation. This makes sense because, as seen in previous work, galaxies and MBHs stop growing once their gas is removed by ram pressure stripping. This means that while the MBHs may spend time away from centre, the galaxies do not have much chance to grow in the absence of MBH accretion, as they do in the field. However, there is evidence of a spread toward lower accreted masses at a similar stellar mass scale as in Romulus25.

\end{itemize}

Taken together, these results predict a population of under-grown MBHs in 
galaxies with $M_{\star}\sim10^{10-11}$ M$_{\odot}$ caused by complex dynamical
histories. This provides a new consideration for causes in scatter in empirical relations. 
Future observations, including gravitational wave measurements, could place constraints on MBH spin, which could encode some of this (lack of) accretion history.
This adds to an increasing
body of work demonstrating the importance of MBH dynamics to their evolution. Modelling
such complex histories is crucial for understanding the diverse histories of MBHs
and their impact on their host galaxies.

\section*{Acknowledgements}

This work used the pynbody \citep{pontzen_pynbody_2013} and Tangos \citep{pontzen_tremmel_tangos_2018} software packages. LAW wishes to acknowledge that this publication has emanated from research conducted with the financial support of Taighde \'Eireann - Research Ireland  under Grant number GOIPG/2025/8041. PN gratefully acknowledges support from the Black Hole Initiative at Harvard University, which is funded by grants from the John Templeton Foundation and the Gordon and Betty Moore Foundation.

\section*{Data Availability}

The Tangos database for both Romulus simulations are available upon email request.

\begin{small}
\bibliographystyle{aa} 
\bibliography{ref}
\end{small}
\end{document}